\documentclass[12pt]{article}
\usepackage{graphicx}
\usepackage{amssymb}
\usepackage{epsfig}
\usepackage{amsmath}
\usepackage{axodraw}
\newcommand{\be}{\begin{equation}}
\newcommand{\ee}{\end{equation}}

\newcommand{\LA}{\langle}
\newcommand{\RA}{\rangle}

\newcommand{\bea}{\begin{eqnarray}}
\newcommand{\eea}{\end{eqnarray}}

\newcommand{\bi}{{\bf i}}
\newcommand{\bj}{{\bf j}}

\newcommand{\Wick}[1]{\;\,\rule[10pt]{.5pt}{2pt}\!\hspace{1.2pt}\rule[12pt]{#1}{.5pt}
\!\hspace{1.1pt}\rule[10pt]{.5pt}{2pt}\hspace{-#1}\hspace{-1.4mm}}

\newcommand{\Wickunder}[1]{\;\,\rule[-6pt]{.5pt}{2pt}\!\hspace{1.2pt}\rule[-6pt]{#1}{.5pt}
\!\hspace{1.1pt}\rule[-6pt]{.5pt}{2pt}\hspace{-#1}\hspace{-1.4mm}}

\DeclareSymbolFont{AMSa}{U}{msa}{m}{n}
\DeclareSymbolFont{AMSb}{U}{msb}{m}{n}
\DeclareMathSymbol{\fieldR}{\mathalpha}{AMSb}{"52}

\newif\ifpdf
\ifx\pdfoutput\uundefined
\pdffalse 
\else
\pdfoutput=1 
\pdftrue
\fi

\begin{document}
\begin{titlepage}
\begin{center}
\hfill {\tt YITP-SB-04-61}\\
\hfill {\tt hep-th/0411171}\\
\vskip 20mm

{\Large {\bf Orbifolding the Twistor String}}

\vskip 10mm

{\bf S.~Giombi, M.~Kulaxizi, R.~Ricci,\\
D.~Robles-Llana, D.~Trancanelli and K.~Zoubos
\footnote{E-mails: {\tt sgiombi,~kulaxizi,~rricci,~daniel,
~dtrancan,~kzoubos@insti.physics.sunysb.edu}}}

\vskip 4mm {\em  C. N. Yang Institute for Theoretical
Physics,}\\
{\em State University of New York at Stony Brook}\\
{\em Stony Brook, NY 11794-3840, USA}\\
[2mm]

\end{center}

\vskip 1in

\begin{center} {\bf ABSTRACT }\end{center}
\begin{quotation}
\noindent

The D-instanton expansion of the topological B-model on the supermanifold $CP^{(3|4)}$ reproduces the perturbative expansion of $\mathcal{N}=4$ Super Yang-Mills theory. In this paper we consider orbifolds in the fermionic directions of $CP^{(3|4)}$. This operation breaks the $SU(4)_R$ $R$-symmetry group, reducing the amount of supersymmetry of the gauge theory. As specific examples we take $\mathcal{N}=1$ and $\mathcal{N}=2$ orbifolds and obtain the corresponding superconformal quiver theories. We discuss the D1 instanton expansion in this context and explicitly compute some amplitudes.

\end{quotation}

\vfill
\flushleft{November 2004}

\end{titlepage}

\eject


\tableofcontents


\section{Introduction}
The conjecture by Witten \cite{witten} that perturbative ${\cal N}=4$ super Yang-Mills can be realized as a D-instanton expansion of the topological B-model with $CP^{(3|4)}$ super-twistor space as target has generated a lot of interest recently. By now tree-level amplitudes are moderately well understood, either directly from the structure of the D-instanton moduli space \cite{roiban}, or from the twistor-inspired field theory procedure of \cite{MHV} \footnote{See \cite{Zhu:2004kr} for subsequent developments and \cite{Khoze:2004ba} for a review.}, which were shown to be related in \cite{gukov}. More recently, some substantial progress has also been achieved in understanding one-loop amplitudes \cite{one-loop}. Interesting alternative proposals to Witten's construction have been put forward in \cite{berkovits}, starting from conventional open strings moving in $CP^{(3|4)}$, \cite{vafa}, where a mirror symmetric A-model version is considered, and \cite{siegel}, where ADHM twistors are considered. Conformal supergravity has been studied in this approach in \cite{bw} and \cite{Ahn:2004yu}, while the possibility of extending the twistor formalism to ordinary gravity has been investigated in \cite{Giombi:2004ix}. Other recent related work can be found in 
\cite{Popov:2004rb}.

Surprising and elegant as Witten's proposal undoubtedly is, it has two obvious shortcomings. The first one is that so far it applies only to maximally supersymmetric gauge theories \footnote{However, examples of twistor-inspired computations of amplitudes in theories with less supersymmetry have appeared in the literature.}; the second is that superconformal invariance is automatically built-in by virtue of the twistor formalism. These features seems to make the original construction unfit for describing more realistic gauge theories. The problem of reducing the model to an $\mathcal{N}=1$ superconformal theory has been recently considered in \cite{Kulaxizi:2004pa}, where a Leigh-Strassler deformation of $\mathcal{N}=4$ SYM was implemented via open/closed couplings. 

In this paper we consider a different extension of Witten's correspondence to a class of ${\cal N}=2$ and ${\cal N}=1$ supersymmetric gauge theories. As we do not know how to relax the requirement of superconformal invariance, natural candidates are the superconformal quiver theories analyzed in \cite{kachru} and \cite{nekrasov}: In this paper we will recover them from twistor strings.

The procedure followed in \cite{kachru} and \cite{nekrasov} was to start with a parent ${\cal N}=4$ super Yang-Mills theory and retrieve the superconformal daughter theories with reduced supersymmetry by orbifolding the $SU(4)_R$ symmetry rotating the supercharges \footnote{Strictly speaking, this is not really an orbifold in the conventional sense of the word, as one is not gauging the discrete symmetry in spacetime.}. These are quiver theories with bifundamental matter fields. In the present case of twistor strings, the $SU(4)_R$ symmetry is part of the isometry group of $CP^{(3|4)}$. Thus, before Penrose transforming, this operation has a natural interpretation as a fermionic orbifold of the twistor string's target space.

Although it is not clear {\it a priori} what the meaning of a fermionic orbifold is, one is immediately tempted to establish a connection with the standard lore about D-branes tranverse to bosonic orbifold singularities \cite{douglas-moore}, and their realization via geometric engineering \cite{geometric engineering}. In the case of ${\cal N}=2$ superconformal theories engineered from type $IIB$ superstrings \footnote{These are necessarily of affine $ADE$ type.}, the moduli space of superconformal couplings is known to admit a duality group whose action is inherited from the S-duality of $IIB$ superstrings \footnote{Alternatively, it can be seen as the affine Weyl group acting on the primitive roots of affine $ADE$ group \cite{Witten:1997sc} \cite{Katz:1997eq} \cite{nekrasov}.}. It would be interesting to identify these moduli spaces in the twistor string theory.

This paper is organized as follows: In Section 2, we briefly recall basic facts about D-branes on  (bosonic) orbifolds. In Section 3, after reviewing the twistor construction of \cite{witten}, we perform the orbifold on the fermionic directions of the super-twistor space, paying particular attention to the role of the $D1$ branes. Consistency will require the introduction of $|\Gamma|$ D1 branes, where $|\Gamma|$ is the order of the orbifold group. In Section 4, we investigate two examples of $\mathcal{N}=1$ and $\mathcal{N}=2$ orbifolds, and explicitly compute some amplitudes. In Section 5, we give the conclusion and some final remarks.


\section{Branes transverse to orbifold singularities}
Placing a set of $D$ branes transverse to a $C^n/\Gamma$ orbifold (where $n=2,3$, and $\Gamma$ is a discrete subgroup of $SU(n)$) gives rise to four dimensional gauge theories with ${\cal N}=2$ or ${\cal N}=1$ supersymmetry living in the brane worldvolume. To obtain their massless spectrum one has to consider the appropriate orbifold action on fields in both the open and closed string sectors. The open string sector contributes the field content of the gauge theories, which can be encoded in quiver diagrams. The closed string sector in turn contributes the necessary moduli which deform the transverse orbifold singularity to a smooth space. We briefly review this construction below, focusing in the case of abelian $\Gamma$ for simplicity.   
 
\subsection{Open string sector}
In the open string sector, ${\Gamma}$ acts on the orbifolded transverse coordinates and on the Chan-Paton factors of open strings in the worldvolume directions. To get conformal theories one needs $\Gamma$ to act on the Chan-Paton factors in (arbitrary number of copies of) the regular representation $\mathcal{R}$ of ${\Gamma}$. Choosing $N$ copies of $\mathcal{R}$ amounts to considering $N|\Gamma|$ $D3$-branes before the projection, where $|\Gamma|$ is the order of $\Gamma$. At this stage one then has $U(N|\Gamma|)$ gauge symmetry in the worldvolume.

In the worldvolume directions $\Gamma$ acts on the open string Chan-Paton factors matrices $\lambda$ only. This action is specified by a matrix ${\gamma}_{\Gamma}{\in}\mathcal{R}$, and the invariant states satisfy
\be
\label{actionp}
{\gamma}_{\Gamma}{\lambda}{\gamma}_{\Gamma}^{-1}={\lambda}.
\ee
Now using elementary group theory the regular representation can be decomposed in irreducible representations as
\be
\mathcal{R}={\oplus}_{\bi}n_{\bi}\mathcal{R}_{\bi}
\ee
where $n_{\bi}={\rm dim}(\mathcal{R}_{\bi})=1$ for abelian $\Gamma$. Acting on ${\lambda}$ as in (\ref{actionp}) this decomposition projects out the fields whose Chan-Paton indices are not connected by one of the irreducible $\mathcal{R}_{\bi}$'s. Taking $N$ copies of $\mathcal{R}$ the gauge group will therefore be broken \footnote{Actually there is a $U(1)$ subgroup acting trivially on the fields. It can be seen as the motion of the center of mass coordinate of the D-branes. Therefore the effective gauge group is $G=F/U(1)$.} to $F={\prod}_{\bi{\in}irreps}U(N)$. Each of these unitary groups with its gauge multiplet has an associated node in the quiver diagram.
 
In the orbifolded directions $\Gamma$ acts on the Chan-Paton matrices through an element of the regular representation $\gamma_\Gamma$, and on space indices through the defining $n$ dimensional representation $G_{\Gamma}^{n{\times}n}$, in such a way that
\be
\label{pro}
G_{\Gamma}^{n{\times}n}({\Psi}(\bi))={\Psi}({\gamma}_{\Gamma}(\bi))
\ee     
${\Psi}(\bi)$ being a label for the $\bi$-th $D3$-brane in the orbifolded transverse space. This means that the action of the space group is correlated with the action on the Chan-Paton factors. The fields surviving the projection (\ref{pro}) can be obtained from the decomposition
\bea
{\rm Hom}(\mathcal{R},G_{\Gamma}^{n\times n}\otimes\mathcal{R})&=&\bigoplus_{\bi,\bj}\left[{\rm Hom}(\mathcal{R}_{\bi},G_{\Gamma}^{n\times n}\otimes\mathcal{R}_{\bj})\otimes {\rm Hom}(C^N,C^N)\right]\\{\nonumber}
&=&\bigoplus_{\bi,\bj}a_{\bi\bj}{\rm Hom}(C^N,C^N)
\eea
where again ${\bi}$ runs over irreducible representations, and $a_{\bi\bj}$ are the Clebsch-Gordan coefficients in the decomposition of the tensor product. Physically these are $a_{\bi\bj}$ chiral multiplets transforming in bifundamental representations as
\be
\label{chirals}
{\oplus}a_{\bi\bj}({\bf{N}},{\bar{\bf{N}}}).
\ee 
The quiver diagram has $a_{\bi\bj}$ oriented links between nodes $\bi$ and $\bj$. For $\mathcal{N}=2$   quivers $a_{\bi\bj}=a_{\bj\bi}$, which makes the links non oriented. Each of them represents an $\mathcal{N}=2$ hypermultiplet.

Finally, for $C^2/\Gamma$ orbifolds one has two non-orbifolded transverse directions. $\Gamma$ acts on these fields as in (\ref{actionp}). They provide the adjoint chiral superfields which together with the gauge multiplets complete the ${\cal N}=2$ vector multiplet. 

\subsection{Closed string sector}
In the closed string sector there are no Chan-Paton factors, and one can follow the ordinary orbifold techniques to find the spectrum. There are, in addition to the usual untwisted sector, $|\Gamma|-1$ twisted sectors which play a crucial role in resolving the singularity. The untwisted sector is just the Kaluza-Klein reduction of the ten dimensional supergravity multiplet on $C^2/\Gamma$ (for ${\cal N}=2$) or on $C^3/\Gamma$ (for ${\cal N}=1$), together with the usual matter multiplets. In the large volume limit the moduli fields from the $|\Gamma|-1$ twisted sectors can be seen to arise by wrapping the various form fields on the exceptional cycles of the blown-up singularity. For ${\cal N}=2$ the blow-up is hyper-K\"ahler, whereas for ${\cal N}=1$ it is only K\"ahler. In the first case there are moduli $b_{\bi}=\int_{S_{\bi}^2}B$ and $\vec \zeta_{\bi}=\int_{S_{\bi}^2}\vec \omega$, where for $IIA$ $B$ is the $NSNS$ B-field \footnote{For $IIB$ one can in addition have the two-form from the $RR$ sector.}, and $\vec \omega$ is the triplet of K\"ahler forms on the blow up. In the K\"ahler case $b_{\bi}=\int_{S_{\bi}^2}B$ and $\zeta_{\bi}=\int_{S_{\bi}^2} \omega$, where $\omega$ is the K\"ahler form. In the first case the combination $b_{\bi}+i\vec \zeta_{\bi}$ encodes (in the large volume limit) the deformations of K\"ahler and complex structure of the resolution. Because of ${\cal N}=2$ supersymmetry the resolution is hyper-K\"ahler, and these two are related by the $SO(3)$ symmetry that rotates $\vec \omega$. In the second case $b_{\bi}+i\zeta_{\bi}$ parameterize the deformations of the complexified K\"ahler structure \footnote{In this case, in addition to $B$ there are further moduli from the $RR$ sector: a hypermultiplet for type $IIB$ or a vector multiplet for type $IIA$.}.

Of course one has in addition to these moduli scalars various $p$-form fields from the twisted sectors. The twisted fields couple to open fields in the brane low effective action via Chern-Simons couplings. In the presence of the orbifold, closed fields from the $k$-th twisted sector $C_{k}$ couple naturally to the $U(1)$ part of the field strength of the $D$-brane whose Chan-Paton factor is twisted by ${\gamma}_{\Gamma}$. Their supersymmetric completion involves terms which couple as Fayet-Iliopoulos parameters in the effective gauge theory on the brane world-volume. When these are non-zero the gauge symmetry is completely broken, and the Higgs branch of the world-volume theory is the resolved transverse space.

Taking into account all the twisted moduli one can write the full stringy quantum volume of the exceptional cycles of the geometry as $V_{\bi}=\left(b_{\bi}^2+|\vec\zeta_{\bi} |^2\right)^{1/2}$ in the ${\cal N}=2$ case and $V_{\bi}=\left(b_{\bi}^2+\zeta_{\bi}^2\right)^{1/2}$ in the ${\cal N}=1$ case. At the orbifold point $\zeta_{\bi}=0$, and one can write the coupling constant of the $\bi$-th gauge group as $1/(g_{YM \, \bi})^2 = V_{\bi}/g_s$, where $g_s$ is the string coupling constant. In type $IIB$ one has also $c_{\bi}=\int_{S^2_{\bi}}B_R$, which plays the role of a theta angle in the gauge theory. One can then write the complexified couplings as $\tau_{\bi}=\theta_{\bi}+i/(g_{YM \, \bi})^2=c_{\bi}+b_{\bi}\tau$, with $\tau=g_s^{-1}$. The S-duality of type $IIB$ superstrings, which acts on $B_{NS}$ and $B_R$ manifests itself as a duality in the moduli space of couplings.   
\section{The orbifold of the twistor string}
\setcounter{equation}{0}

\subsection{Review of the topological B-model on $CP^{(3|4)}$}

We start by reviewing the model proposed by Witten \cite{witten} to describe perturbative $\mathcal{N}=4$ SYM: This is  a topological B-model on the supermanifold $CP^{(3|4)}$. The local bosonic and fermionic coordinates on this space are $(Z^I,\psi^A)$, with $I=1, \ldots , 4$ and $A=1, \ldots , 4$. They are subject to the identification $(Z^I,\psi^A)\sim(tZ^I,t\psi^A)$, with $t\in C^*$. The $Z^I$ cannot be all zero. Equivalently, this supermanifold can be defined as the sublocus of $C^{(4|4)}$ 
\bea
\left(\sum_{I}|Z^I|^2+\sum_{A}|\psi^A|^2=r\right)/U(1)
\eea
where $U(1)$ is a phase transformation acting as $(Z^I, \psi^A)\rightarrow e^{i\alpha}(Z^I, \psi^A)$.
This space is a super Calabi-Yau with holomorphic volume form  
\bea
\Omega=\frac{1}{(4!)^2}\epsilon_{IJKL}\epsilon_{ABCD}Z^I dZ^J dZ^K dZ^L d\psi^A d\psi^B d\psi^C d\psi^D.\label{Om}
\eea
In this model the self-dual part of $\mathcal{N}=4$ SYM with gauge group $U(N)$ is reproduced by the world volume action of a stack of $N$ D5 branes. These are almost space-filling branes placed at $\bar{\psi}_A=0$.
The world volume action is a holomorphic Chern-Simons 
\bea
\mathcal{S}=\int_{D5}\Omega\wedge Tr \left(\mathcal{A}\bar{\partial}\mathcal{A}+\frac{2}{3}\mathcal{A}\wedge\mathcal{A}\wedge\mathcal{A}\right)
\eea
where $\mathcal{A}=\mathcal{A}_{\bar{I}}d\bar{Z}^{\bar{I}}$ is a $(0,1)$ form with values in the adjoint representation of $U(N)$. The superfield expansion reads
\bea
\mathcal{A}(z,\bar{z},\psi)&=&A(z,\bar{z})+\psi^A\chi_A(z,\bar{z})+\frac{1}{2}\psi^A\psi^B\phi_{AB}(z,\bar{z}) \nonumber \\
&+&\frac{1}{3!}\epsilon_{ABCD}\psi^A\psi^B\psi^C\tilde{\chi}^D(z,\bar{z})+\frac{1}{4!}\epsilon_{ABCD}\psi^A\psi^B\psi^C\psi^D G(z,\bar{z}).\label{A}
\eea
The components of $\mathcal{A}$ are charged under the symmetry 
\bea S: \psi^A \rightarrow e^{i\beta}\psi^A\label{Ssym}
\eea as $S(A,\chi,\phi,\tilde{\chi},G)=(0,-1,-2,-3,-4)$. One can integrate out the fermionic coordinates to get
\bea
\mathcal{S}&=&\int_{CP^3}\Omega'\wedge Tr [G\wedge(\bar{\partial}A+A\wedge A)+\tilde{\chi}^A\wedge \bar{D}\chi_A\nonumber \\
&+&\frac{1}{4}\epsilon^{ABCD}\phi_{AB}\wedge \bar{D}\phi_{CD}+\frac{1}{2}\epsilon^{ABCD}\chi_A\wedge\chi_B\wedge\phi_{CD}]\label{action}
\eea
where $\Omega'$ is the bosonic reduction of $\Omega$.
After performing a Penrose transform \cite{penrose}, (\ref{A}) yields the field content of the $\mathcal{N}=4$ vector multiplet, whereas (\ref{action}) reproduces the self-dual truncation of the $\mathcal{N}=4$ SYM action \cite{siegel2}. 


The non self-dual completion of the theory is obtained by introducing D1 instanton corrections. These are holomorphic maps from  $CP^{1}$ to $CP^{3|4}$ and the instanton number corresponds to the degree of the map. We will discuss later in more detail the role played by these D1 branes.

\subsection{The orbifold}

The procedure reviewed in Section 2 can be applied to the twistor string of \cite{witten} in order to reduce the ${\cal N}=4$ supersymmetry. 
The homogeneous coordinates of $CP^{3|4}$ provide a linear realization of $PSU(4|4)$, which is the ${\cal N}=4$ superconformal group. It is therefore natural to use twistors to study conformal theories. To reduce supersymmetry, we can  orbifold the fermionic directions of the super-twistor space. Physically, this amounts to orbifolding the $SU(4)_R$ $R$-symmetry, which is the Fermi-Fermi subgroup of $PSU(4|4)$. As reviewed above, in the twistor theory of \cite{witten} a set of D5 branes is placed at $\bar{\psi}_A=0$. In analogy with the conventional case, one possible interpretation is to view the orbifold as acting in the $\bar{\psi}_A$ directions, which are transverse to the D5 branes. This induces an action on the $\psi^A$ which will be the one considered in the following. 

Explicitly, we choose an action of the orbifold $\Gamma \in Z_k$ under which the fermionic coordinates transform as
\bea
\psi^A\rightarrow e^{2\pi i a_A/k}\psi^A\label{psiorb}
\eea
with the condition on the charges $\sum_{A}a_A=0$ (mod $k$), so that $\Gamma \in SU(4)_R$. The holomorphic volume form (\ref{Om}) is invariant under (\ref{psiorb}): This implies that the super-orbifold is still Calabi-Yau. This is crucial for the consistency of the B-model \cite{WittenSig}. 

\begin{figure}
\begin{center}
\includegraphics[width=40mm]{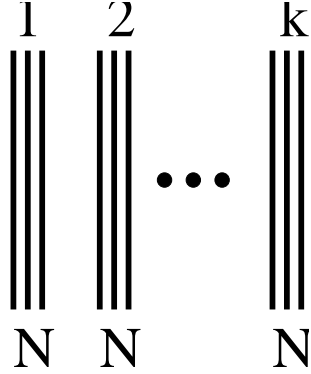}
\caption{Regrouping of the D5 branes under the orbifold action. Note that the $k$ stacks of branes are actually coincident at the point $\bar{\psi}_A=0$.}\label{branes}
\end{center}
\end{figure}

We consider a stack of $kN$ D5 branes in the covering space. The orbifold action regroups the branes in $k$ stacks of $N$ branes each, as shown in Figure \ref{branes}. It is thus convenient to decompose the $U(kN)$ adjoint index into $A^a_{\;\; b}=A^{I\bi}_{\;\; J\bj}$, where $I,J=1,\ldots,N$ label the brane within a stack  and $\bi,\bj=1,\ldots,k$ label the stacks. 

An explicit representation of $\Gamma$ is 
\bea
\mathcal{R}=\left(\begin{array}{cccc}
                    r_1 & 0       & \cdots    & 0      \\
                     0  & r_2     & \cdots    & 0      \\
                 \vdots & \vdots  & \ddots    & \vdots \\
                    0   &  0      & \cdots    & r_k    \\
\end{array}\right)_{kN\times kN}\label{svista}
\eea
where $r_\bi=e^{2\pi i \bi/k}$ is a $N\times N$ matrix acting on the $\bi$-th node of the associated quiver. The orbifold projection is enforced by requiring invariance of the components of the superfield $\mathcal{A}$ under the action of $\Gamma$.
$R$-symmetry invariance of the superfield implies that (\ref{psiorb}) induces a conjugate transformation on the fermionic indices of the components. Combined with the action on the Chan-Paton factors given by (\ref{svista}), this gives the following orbifold action on a generic component (with $n=0, \ldots, 4$ fermionic indices) \footnote{In this notation, for instance, $\Phi_{A_1A_2A_3A_4}=\frac{1}{4!}\epsilon_{A_1A_2A_3A_4}G$, where $G$ is the highest component of $\mathcal{A}$.}  
\bea
(\Phi_{A_1,\ldots, A_n})^{I\bi}_{\;\; J\bj} \rightarrow e^{2\pi i(\bi-\bj-a_{A_1}-\ldots -a_{A_{n}})/k}(\Phi_{A_1,\ldots, A_n})^{I\bi}_{\;\; J\bj}.
\eea
For example, the lowest component of (\ref{A}), which physically represents the positive helicity gluon, transforms as 
\bea
A^{I\bi}_{\;\;J\bj}\rightarrow e^{2\pi i(\bi-\bj)/k}A^{I\bi}_{\;\;J\bj}.
\eea
Invariance requires $\bi=\bj$, so that the gauge group is broken to
\bea
U(kN)\rightarrow[U(N)]^k.
\eea
Similarly, the positive helicity gluino $\chi_A$ transforms as 
\bea
(\chi_A)^{I\bi}_{\;\;J\bj} \rightarrow e^{2\pi i (\bi-\bj-a_A)/k}(\chi_A)^{I\bi}_{\;\;J\bj}
\eea
so that in this case one needs to enforce $\bi=\bj+a_A$. Depending on the value of the charge $a_A$, the field $\chi_A$ becomes either a gaugino or a bi-fundamental quark.  One proceeds analogously with the other components of $\mathcal{A}$. One can picture the field content in a quiver 
diagram with $k$ nodes corresponding to the $k$ gauge groups and bi-fundamental matter as lines connecting pairs of nodes.   

The choice of the discrete group one quotients by determines the amount of supersymmetry preserved by the orbifold. For generic $\Gamma$ the supersymmetry is completely broken, while for $\Gamma\in Z_k \subset  SU(2)_R$ and $\Gamma\in Z_k \subset  SU(3)_R$ one has respectively $\mathcal{N}=2$ and $\mathcal{N}=1$ \cite{kachru}.

So far we have only focused on the D5 brane sector. In the next Section we will tackle the problem of orbifolding the D1 instantons.

\subsection{D1 branes}

As already remarked, the holomorphic Chern-Simons action on $CP^{3|4}$ only reproduces the self-dual truncation of $\mathcal{N}=4$ SYM. A non-perturbative correction to the B-model is needed in order to recover the non self-dual part of the gauge theory. These new non-perturbative degrees of freedom are D1 branes wrapped on holomorphic cycles inside the supermanifold. These branes are D-instantons whose instanton number is given by the degree $d$ of the map. The simplest case \footnote{We will only consider tree-level scattering amplitudes. Amplitudes with $l$ loops receive contributions also from curves of genus $g \leq l$.} is genus $g=0$ and $d=1$. This is the only example we discuss in this paper. The explicit map is
\bea
& &\mu_{\dot a}+x_{a \dot a}\lambda^{a}=0\nonumber\\
& &\psi^A+\theta^{A}_{a}\lambda^a=0.
\eea
Here the coordinates of $CP^3$ are decomposed as $Z^I=(\lambda^a,\mu_{\dot a})$ and $x_{a\dot a}, \theta^A_a$ are the moduli of the embedding. In \cite{witten} the D1 instanton is initially placed at $\psi^A=0$ and the dependence on the fermionic coordinates is then restored through an integration over the moduli space.

Tree-level gauge theory scattering amplitudes are computed by considering the effective action for D1-D5 strings
\bea
I_{D1-D5}=\int_{D1} dz \,\beta \bar{D} \alpha
\label{I}
\eea
where $\alpha$ and $\beta$ are fermions which carry respectively fundamental and anti-fundamental gauge group indices. They correspond to strings stretching from the D1 to the D5 brane and viceversa. The covariant derivative is $\bar{D}=\bar{\partial}+\mathcal{A}$. 
The action (\ref{I}) contains the interaction term 
\bea
\Delta I_{D1-D5}=\int_{D1} Tr J\mathcal{A}=\int_{D1} J^{a}_{\;\; b}\mathcal{A}^{b}_{\;\; a}
\label{inter}
\eea
where $J^a_{\;\; b}=\alpha^a \beta_b$. Scattering amplitudes are obtained by taking correlation functions of the currents $J$'s in the background of the superfield $\mathcal{A}$ and integrating them over the moduli space of a D1 instanton of appropriate degree. This degree is determined by the sum over the $S$ symmetry (\ref{Ssym}) charges of the $n$ external states
\bea
d=-\frac{1}{4}\sum_{i=1}^n S_i-1.
\eea 
In the particular case of external gluons, this corresponds to $d=q-1$, where $q$ is the number of negative helicity gluons. Explicitly, when $d=1$ one has for the $n$-point scattering amplitude
\bea
A_n=\int d^8 \theta\, w_1\ldots w_n \LA J^{a_1}_{\;\; 1}\ldots J^{a_n}_{\;\; n} \RA\label{ampl}
\eea
where the $w_i$ are the wave functions of the external states and are given essentially by the coefficient of that state in the superfield expansion. We stress that the measure in (\ref{ampl}) is going to be invariant under the orbifold action.

We now discuss the issue of how the coupling constant may arise in the theory. From a four dimensional field theoretical point of view, the completion of the self-dual Yang-Mills action $I_{SD}=\int d^4x \, Tr (GF)$ is given by
\bea
I_{YM}=\int d^4x \, Tr \left(GF-\frac{g^2_{YM}}{2}G^2\right).
\eea 
Therefore, in the topological B-model one expects the coupling constant to originate from the D1 instanton expansion. This is rather surprising since now the YM perturbative coupling seems to come from non-perturbative sectors of the theory. 
One natural way to introduce a free parameter in the amplitude (\ref{ampl}) is to weigh it by a factor $(e^{-I_{D1}})^d$, where $I_{D1}$ is the action for a D1 instanton of degree $d=1$. This has been already remarked in \cite{witten}. Another way to achieve this is to consider the coupling of the D1 to the closed sector of the B-model. This was first realized in \cite{bw} where a new field $b$, a (1,1) form in twistor space, is introduced. It has a minimal coupling to the D1 worldvolume
\bea
I_b=\int_{D1}b_{I\bar{J}}\, dZ^{I}\wedge d\bar{Z}^{\bar J}.\label{bfield}
\eea
This field is not present in the perturbative analysis of the B-model. The necessity of it was also recently rediscussed in \cite{VafaM} as a non-perturbative correction to Kodaira-Spencer theory \cite{BCOV}. For a D1 sitting at $(x,\theta)$ in the moduli space the coupling (\ref{bfield}) directly defines the conformal supergravity superfield $\mathcal{W}(x,\theta)=\int_{D1_{(x,\theta)}}b$. The lowest component can be interpreted as a dilaton $C$. As a consequence of the coupling (\ref{bfield}), in a vacuum with expectation value $\langle C \rangle=c$, an amplitude will be weighted by a factor $(e^{-c})^d$. This is reminiscent of ordinary string theory where the coupling constant comes from the dilaton expectation value. 

In summary, we assume the contribution of the D1 instanton to an amplitude to be equal to $(g^2)^d$, where $g^2$ might come from $e^{-I_{D1}}$ or $e^{-c}$.

To get the standard normalization of the scattering amplitudes, one also needs to rescale each component of the superfield $\mathcal{A}$ by a factor $g^{1+\frac{1}{2}S}$, where $S$ is the charge under the symmetry (\ref{Ssym}): For instance, $A$ goes to $gA$, $\chi$ to $\sqrt{g}\chi$, and so on. In the end, the overall coupling constant in a tree-level $n$-point amplitude is  
\bea
\left( \prod_{i=1}^n g^{1+\frac{1}{2}S_i}\right) (g^2)^{-\frac{1}{4}\sum_{i}S_i-1}=g^{n-2}.
\eea

\begin{figure}
\begin{center}
\includegraphics[width=150mm]{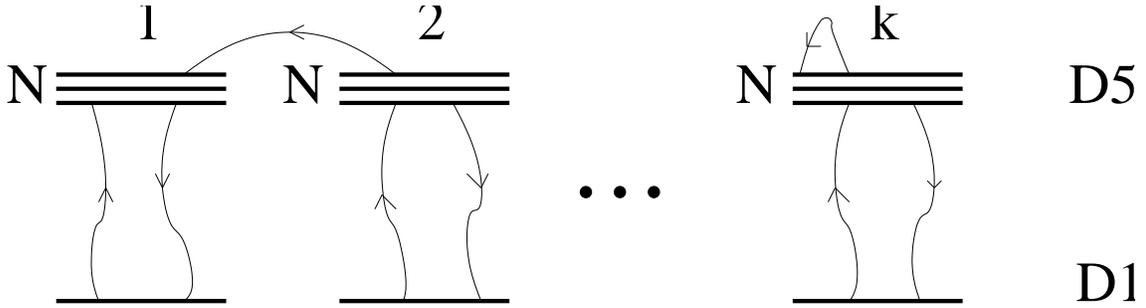}
\caption{The stacks of D5 and D1 branes. An interaction between the first and the second stack is depicted.}\label{d1d5}
\end{center}
\end{figure}
\begin{figure}
\begin{center}
\includegraphics[width=90mm]{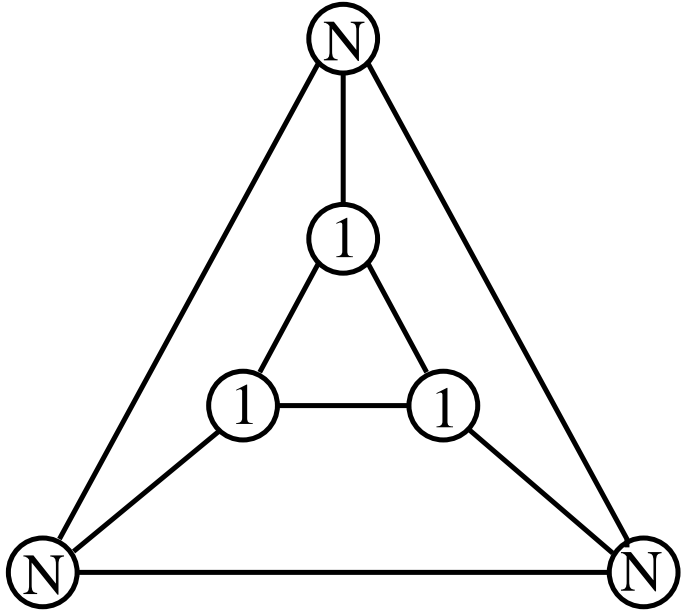}
\caption{The quiver for the D1-D5 brane system in the case $k=3$.}\label{Q15}
\end{center}
\end{figure}

We now proceed to the analysis of the orbifold action on the D1 instanton sector. For this action to be faithful on the Chan-Paton factors of the D1's, we need to start with $k$ D1 branes. To begin with, we locate them at $\psi^A=0$. As in \cite{witten}, the fermionic dependence will be restored in the end through integration over the moduli space. Considering $k$ D1 branes, the effective action (\ref{I}) gets changed into
\bea
I_{D1-D5}=\int_{D1} dz \,(\beta^r_{I\bi}\bar{\partial}\alpha^{I\bi}_r+\beta^r_{I\bi}\mathcal{B}_{r}^{\;\;s}\alpha^{I\bi}_s+\beta^r_{I\bi}\mathcal{A}^{I\bi}_{\;\; J\bj}\alpha_r^{J\bj})
\label{Imod}
\eea
where $r=1,\ldots, k$ is a $U(k)$ index which labels the D1's. For instance, $\alpha^{I\bi}_r$ is a string stretching from the $r$-th D1 brane to the $I$-th D5 brane inside the $\bi$-th stack. In (\ref{Imod}) $\mathcal{B}_{r}^{\;\;s}$ is the $U(k)$ gauge field on the world-volume of the D1 branes.
The action of the orbifold breaks $U(k)\rightarrow [U(1)]^k$. The D1-D5 strings $\alpha$ and $\beta$ transform as 
\bea
& &\alpha_r^{I\bi}\rightarrow e^{2\pi i (\bi-r)/k}\alpha_r^{I\bi}\nonumber \\
& &\beta^r_{I\bi}\rightarrow e^{2\pi i (r-\bi)/k}\beta^r_{I\bi}.\label{D1D5}
\eea
Invariance under the orbifold action requires $\bi=r$. This implies that the D1-D5 strings only stretch between the $\bi$-th D1 brane and the D5 branes in the $\bi$-th stack. This is shown in Figure \ref{d1d5} and, in quiver language, for the specific example of $k=3$, in Figure \ref{Q15}. 

The $U(1)$ fields living on the D1 branes and the bi-fundamental matter connecting them will not be considered in the following, although they are depicted in Figure \ref{Q15}. The $U(1)$ bundles over curves of genus $g < 2$ do not have moduli as remarked in \cite{witten} and do not play a role in the computation of amplitudes. Further, it seems natural to neglect the bi-fundamental fields since in general, when the branes move away from $\psi^A=0$, they should correspond to massive states. 

The stack of $k$ D1 branes can move away from the orbifold fixed point as one full regular brane. In the covering space, the $k$ D1 branes are located in points related by the $\Gamma$ action in the orbifold directions, whereas they coincide in the others. In particular, they coincide along the bosonic subspace and therefore the bosonic world-volume is the same for all of them. Since the branes cannot move independently we have only one set of moduli $(x,\theta)$ for the whole system. However this is not the complete story. We do not fully explore the richness of the orbifold construction: If the branes were coincident at the fixed point of the fermionic coordinates, then there would be no constraint on their motion along the remaining directions and one would have an extended moduli space $\{(x_r,\theta_r)\}$ with $k$ sets of parameters. We will not study this explicitly but we will limit ourselves to some comments. Having $k$ independent D1 branes allows one to consider $k$
independent minimal couplings (\ref{bfield}) to the $b$ field. This might be worth studying because it could provide a mechanism to generate $k$ independent coupling constants, one at each node.
Since in the usual case the moduli space of couplings is related to twisted sectors, it will be valuable to clarify this issue further by studying the closed sector of the B-model and check whether it contains twisted states.

\section{Explicit examples of orbifolds}
\setcounter{equation}{0}

\subsection{An $\mathcal{N}=1$ orbifold}

\begin{figure}
\begin{center}
\includegraphics[width=70mm]{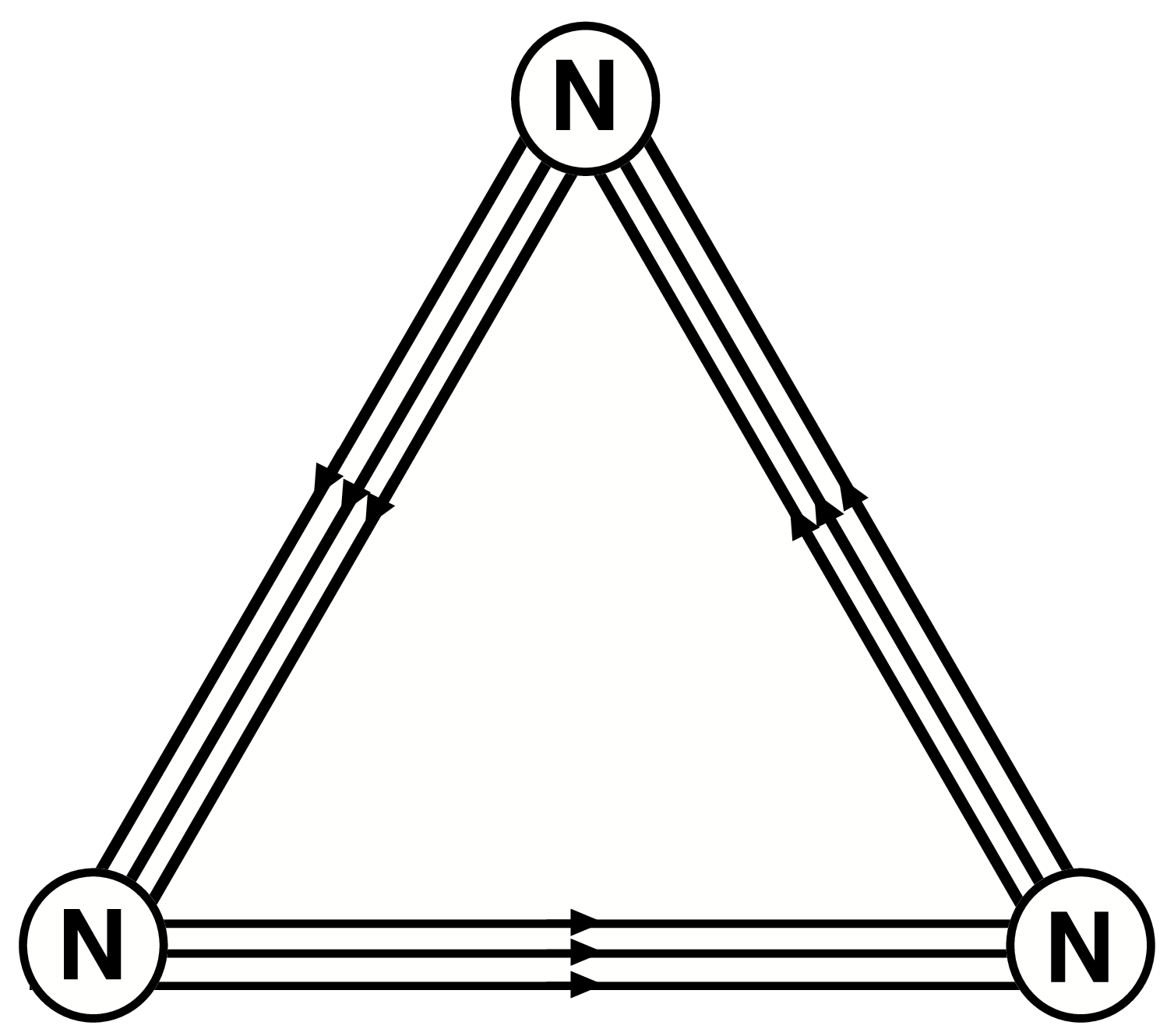}
\caption{The $\mathcal{N}=1$ quiver for $k=3$ and $a_A=(1,1,1,0)$.}\label{fig2}
\end{center}
\end{figure}

We start by considering the case  $\Gamma\in Z_k \subset  SU(3)_R$. Then $SU(4)_R$ is broken into $U(1)_R$, yielding $\mathcal{N}=1$. We consider a particularly simple example, in which $k=3$ and $a_A=(1,1,1,0)$, see (\ref{psiorb}). The gauge group is decomposed into $[U(N)]^3$ and the corresponding quiver diagram is depicted in Figure \ref{fig2}.

The gauge sector contains three $\mathcal{N}=1$ vector multiplets
\bea
A_{\bi},\;\;\;G_{\bi},\;\;\;\lambda_{\bi},\;\;\;\tilde{\lambda}_{\bi}
\eea
where $A_{\bi}\equiv A^{I\bi}_{\;\;J\bi}$, $G_{\bi}\equiv G^{I\bi}_{\;\;J\bi}$, $\lambda_{\bi}^{a}\equiv (\chi_{4})^{I\bi}_{\;\;J\bi}$, $\tilde{\lambda}_{\bi}^{a}\equiv (\tilde{\chi}_{4})^{I\bi}_{\;\;J\bi}$. The index $\bi=1,2,3$ labels the nodes of the quiver. 

On the other hand, the matter sector consists of three $\mathcal{N}=1$ chiral multiplets for each pair of nodes 
\bea
q^{\mu}_{\bi,{\bf i+1}}, \;\;\;\tilde{q}^{\mu}_{{\bf i+1},\bi}, \;\;\; \phi^{\mu}_{\bi,{\bf i+1}}, \;\;\; \tilde{\phi}^{\mu}_{{\bf i+1},\bi}
\eea
where now the index $\mu$ runs from 1 to 3. Here a subscript $\bi,\bj$ indicates that the field has fundamental index in the $\bi$-th node and anti-fundamental in the $\bj$-th node. 
The quarks $q^{\mu}_{\bi,{\bf i+1}}$  and the anti-quarks $\tilde{q}^{\mu}_{{\bf i+1},\bi}$ come from $(\chi^{\mu})^{I\bi}_{\;\;J,{\bf i+1}}$ and $(\tilde{\chi}^{\mu})^{I,{\bf i+1}}_{\;\;J\bi}$. The scalars $\phi^{\mu}_{\bi,{\bf i+1}}$ and $\tilde{\phi}^{\mu}_{{\bf i+1},\bi}$ come from $(\phi_{\mu 4})^{I\bi}_{\;\;J,{\bf i+1}}$ and $\epsilon^{\mu\nu\rho}(\phi_{\nu\rho})^{I,{\bf i+1}}_{\;\;J\bi}$.
The gauge theory with this field content is superconformal \cite{kachru} \cite{nekrasov}. 

\begin{figure}
\begin{center}
\includegraphics[width=130mm]{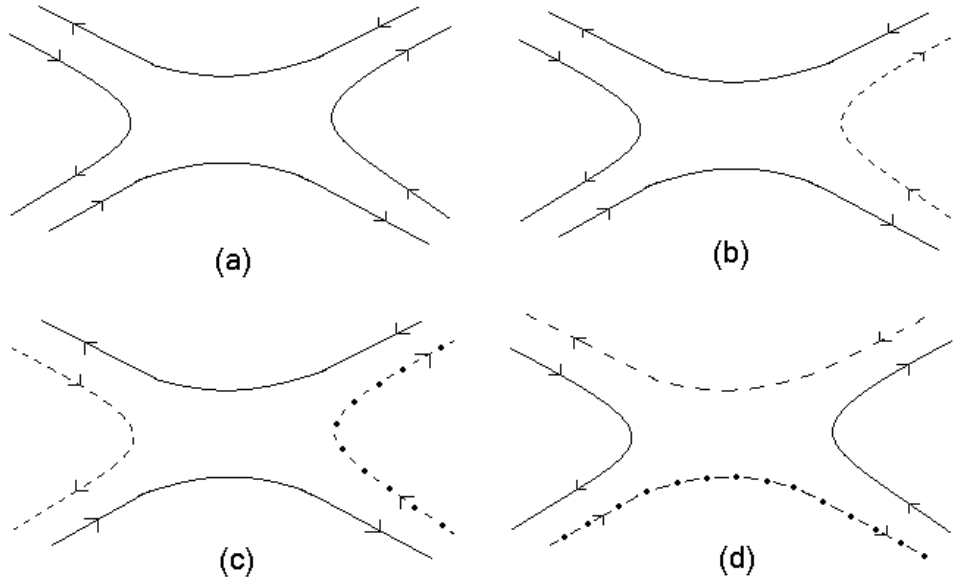}
\caption{The double line notation: (a) scattering of four adjoint fields belonging to the same node; (b) scattering of two adjoint and two bi-fundamental fields; (c) scattering of four bi-fundamental fields with intermediate adjoint field; (d) same as in (c) but with intermediate bi-fundamental field.}\label{doub}
\end{center}
\end{figure}

In terms of these fields the action (\ref{action}) becomes
\bea
\mathcal{S}=\sum_{\bi=1}^3\int_{CP^3}\Omega' &\wedge & Tr [G_{\bi}\wedge(\bar{\partial}A_{\bi}+A_{\bi}\wedge A_{\bi})+\tilde{\lambda}_{\bi}\wedge \bar{D}_{\bi}\lambda_{\bi}\nonumber \\
&+&
\tilde{q}_{\mu \, {\bf i+1},\bi}\wedge \bar{D}_{\bi}\, q^{\mu}_{\bi,{\bf i+1}}+\tilde{\phi}_{\mu \, {\bf i+1},\bi}\wedge \bar{D}_{\bi}\, \phi^{\mu}_{\bi,{\bf i+1}}\nonumber \\
&+& \epsilon_{\mu\nu\rho}\, q^{\mu}_{\bi,{\bf i+1}}
\wedge q^{\nu}_{{\bf i+1},{\bf i+2}}\wedge \phi^{\rho}_{{\bf i+2},\bi}+\lambda_{\bi}\wedge q^{\mu}_{\bi,{\bf i+1}}\wedge \tilde{\phi}_{\mu \, {\bf i+1},\bi}].
\eea
The interaction term (\ref{inter}) becomes after the orbifold projection
\bea
\Delta I_{D1-D5}&=&\int_{D1} Tr J\mathcal{A}=\int_{D1} J^{I\bi}_{\;\; J\bj}\mathcal{A}^{J\bj}_{\;\; I\bi}\rightarrow \nonumber \\
& \rightarrow &\sum_{\bi=1}^3 \int_{D1} Tr [J_{\bi} A_{\bi}+ \psi^4 J_{\bi} \lambda_{\bi} + \psi^{\mu} J_{{\bf i+1},{\bf i}}{q}_{\mu \, {\bf i},{\bf i+1}}\nonumber \\
&+& \frac{1}{2}\epsilon_{\mu\nu\rho}\psi^{\mu}\psi^{\nu}J_{\bi,{\bf i+1}}\tilde{\phi}^{\rho}_{{\bf i+1},\bi}+\psi^{\mu}\psi^4J_{{\bf i+1},\bi}\phi_{\mu \, \bi,{\bf i+1}}\nonumber \\
&+& \frac{1}{3!}\epsilon_{\mu\nu\rho}\psi^{\mu}\psi^{\nu}\psi^{\rho}J_{\bi}\tilde{\lambda}_{\bi}+\frac{1}{2}\epsilon_{\mu\nu\rho}\psi^{\mu}\psi^{\nu}\psi^{4}J_{\bi,{\bf i+1}}\tilde{q}^{\rho}_{{\bf i+1},\bi}
+\psi^1\psi^2\psi^3\psi^4 J_{\bi} G_{\bi}]
\eea
with $J_{\bi}\equiv J^{J\bi}_{\;\; I\bi}=\alpha^{J\bi}\beta_{I\bi}$ and $J_{\bi,{\bf i+1}}\equiv J^{J,{\bf i+1}}_{\;\; I\bi}=\alpha^{J,{\bf i+1}}\beta_{I\bi}$. A convenient way to keep track of the group theory factors is to use a double line notation, where one assigns a different type of oriented line to the fundamental index of each node. For instance, an adjoint field is represented by two lines of the same type and opposite orientation, while a bi-fundamental field has two lines of different type and opposite orienation. For example, in the $\mathcal{N}=1$ case discussed here there are three types of lines corresponding to the three nodes of the quiver  in Figure \ref{fig2}. Some examples are shown in Figure \ref{doub}. 


\subsubsection{MHV Amplitudes for the $\mathcal{N}=1$ orbifold}

In order to calculate MHV amplitudes in these theories we need to follow the general prescription given in \cite{witten} and \cite{Nair}. This prescription is applicable also in this case since we do not allow the $k$ D1 branes to move independently, as already discussed.
Saturation of the fermionic degrees of freedom requires eight $\theta$'s. We consequently have as many different MHV amplitudes, as there are possible 
products of terms from the superfield expansion giving eight $\theta$'s. The denominator of these analytic amplitudes is provided by the current 
correlation functions. The latters also provide the appropriate group structure.
Note that here we should be a little bit more careful than usual, since part of the gauge theory trace is implicit in the summation of the indices $\bi$
which belong to the fundamental representation. We should therefore make sure that we consider only meaningful products of currents, that 
correspond to single trace terms for each MHV analytic amplitude, for instance products like $J_{\bi}J_{\bi,{\bf i+1}}J_{{\bf i+1},\bi}J_{\bi}$ for a four point amplitude 
of the form $(\lambda\, \tilde{q} q\, \tilde{\lambda})$. The possible group theory contractions 
are easily estabilished by drawing the diagrams in double-line notation.

It is now straightforward to proceed to the computation of specific amplitudes of interest. One could rewrite the $\mathcal{N}=4$ superfield expansion (\ref{A}) in $SU(3)\times U(1)_R$ notation which is manifestly $\mathcal{N}=1$ invariant. Recalling that the momentum structure of the amplitudes is solely determined by the form of this expansion, we deduce that analytic amplitudes in the $\mathcal{N}=1$ orbifold theory bear an identical spinor product structure to the ones of $\mathcal{N}=4$ SYM in $SU(3)\times U(1)_R$ notation.
This is in complete accordance with field theoretical considerations, since the Lagrangian description of these theories is identical apart from their group structure. We will make this point clearer with several examples. Note, however, that in what follows we will omit group indices \footnote{As usual we strip out the gauge group theory factor.} and coupling constants, since we stay at the point in the moduli space where all the gauge couplings are equal. 

\paragraph{Example 1: Amplitudes $(A\ldots A\,G\,G)$, $(A\ldots A\,G\,\lambda\,\tilde{\lambda})$, and $(A\ldots A\, G\, q\,\tilde{q})$}

These amplitudes have been extensively considered in the literature (for instance, see \cite{ParkesMangano}). 
As a first trivial check of the above, we compute the standard four gluon amplitude $(A_{\bi}A_{\bi}G_{\bi}G_{\bi})$, with $\bi=1,2,3$. Following the usual
prescription, and using (\ref{A}), we have
\bea
A^{(AAGG)}=\int d^8\theta 
(\psi^1_3\psi^2_3\psi^3_3\psi^4_3)(\psi^1_4\psi^2_4\psi^3_4\psi^4_4)
\frac{1}{\LA 12\RA\LA 23\RA\LA 34\RA\LA 41\RA}=\frac{\LA 34\RA^4}{\LA12\RA\LA23\RA\LA34\RA\LA41\RA}.
\eea
This is the familiar formula for MHV scattering in $\mathcal{N}=4$ SYM. 
In the same way, one can compute amplitudes of the type
$(A\ldots A\,G\,G)$, $(A\dots A\,G\,\lambda\,\tilde{\lambda})$, $(A\dots A\,G\,q\,\tilde{q})$, and $(A\ldots A\,\lambda\,\tilde{\lambda}\,q\,\tilde{q})$.


\paragraph{Example 2: Amplitudes $(q\,q\,\tilde{q}\,\tilde{q})$ and  $(\lambda\, q\,\tilde{q}\,\tilde{\lambda})$}

These amplitudes have, as previously mentioned, the same spinor product structure $\mathcal{N}=4$ SYM has. Yet, they are far more interesting cases to study. The reason is that they consist of two subamplitudes, shown in the case $(q\,q\,\tilde{q}\,\tilde{q})$ in Figure 6.  
They depend on both gluon and scalar particle exchange.
Let us now concentrate on $(q^{\rho}_{{\bf i-1},\bi}\,q^{\sigma}_{\bi,{\bf i+1}}\,\tilde{q}^{\kappa}_{{\bf i+1},\bi}\,\tilde{q}^{\lambda}_{\bi,{\bf i-1}})$.    
The other case $(\lambda\, q\,\tilde{q}\,\tilde{\lambda})$ can be computed in a similar manner. The two subamplitudes in Figure 6 correspond in double line notation to diagrams (d) and (c) in Figure \ref{doub} 
\bea\label{qfouramp}
A^{(qq\tilde{q}\tilde{q})}=\int d^8\theta \frac{1}{4}\psi^{\rho}_1\psi{^\sigma}_2(\psi^4_3\psi^{\mu}_3\psi^{\nu}_3)(\psi^4_4\psi^{\pi}_4\psi^{\tau}_4)
\epsilon_{\mu\nu\kappa}\epsilon_{\pi\tau\lambda}\frac{1}{\LA 12\RA\LA23\RA\LA34\RA\LA41\RA}.
\eea 
Integration over the $\theta^4$ is straightforward and yields $\LA 34\RA$. 
Then, we must sum over all possible contractions of momenta upon integration over the fermionic part of the space. There are three distinct ones
\bea \label{qfourabc}
\begin{split}
(a)&  \int d^6\, \theta 
\Wick{4mm}
\psi^{\rho}_1\psi^{\sigma}_2
\Wick{10mm}
\psi^{\mu}_3
\Wickunder{10mm}
\psi^{\nu}_3\psi^{\pi}_4\psi^{\tau}_4 
+\{ ^{\mu\leftrightarrow \nu}_{\!\,\, \pi\leftrightarrow \tau} \}=
\delta^{\rho\sigma}(\delta^{\mu\tau}\delta^{\nu\pi}-\delta^{\mu\pi}\delta^{\nu\tau})\LA 12\RA\LA34\RA^2\\ \\
(b)&  \int d^6\, \theta 
\Wick{24mm}
\psi^{\rho}_1
\Wickunder{4mm}
\psi^{\sigma}_2\psi^{\mu}_3
\Wickunder{4.5mm}
\psi^{\nu}_3\psi^{\pi}_4\psi^{\tau}_4 
+\{ ^{\mu\leftrightarrow \nu}_{\!\,\, \pi\leftrightarrow \tau} \}
=\left(\delta^{\rho\sigma}(\delta^{\mu\pi}\delta^{\nu\tau}-\delta^{\mu\tau}\delta^{\nu\pi})-\epsilon^{\rho\mu\nu}\epsilon^{\sigma\pi\tau}\right)\LA 23\RA\LA 34\RA\LA 41\RA\\ \\
(c)&  \int d^6 \, \theta 
\Wick{9mm}
\psi^{\rho}_1
\Wickunder{14.5mm}
\psi^{\sigma}_2\psi^{\mu}_3
\Wick{10mm}
\psi^{\nu}_3\psi^{\pi}_4\psi^{\tau}_4 
+\{ ^{\mu\leftrightarrow \nu}_{\!\,\, \pi\leftrightarrow \tau} \}
=\left(\delta^{\rho\sigma}(\delta^{\mu\pi}\delta^{\nu\tau}-\delta^{\mu\tau}\delta^{\nu\pi})-\epsilon^{\rho\pi\tau}\epsilon^{\sigma\mu\nu}\right)\LA 13\RA\LA 24\RA\LA 34\RA
\end{split}
\eea
The three spinor product structures in (\ref{qfourabc}), are related through the Schouten Identity 
\bea \label{Schouten}
\langle pq\rangle\langle rs \rangle+\langle qr\rangle\langle ps \rangle+\langle rp\rangle\langle qs \rangle=0.
\eea
Use of this identity reveals the two independent structures that we were expecting. Explicitly, we have
\bea \label{twofactors}
\int d^6 \, \theta\psi^{\rho}_1\psi^{\sigma}_2\psi^{\mu}_3\psi^{\nu}_3\psi^{\pi}_4\psi^{\tau}_4=
-\epsilon^{\rho\pi\tau}\epsilon^{\sigma\mu\nu}\LA 12\RA\LA 34\RA^2+(\epsilon^{\rho\pi\tau}\epsilon^{\sigma\mu\nu}-\epsilon^{\rho\mu\nu}\epsilon^{\sigma\pi\tau})\LA 23\RA\LA 34\RA\LA 41\RA.
\eea
Inserting (\ref{twofactors}) into (\ref{qfouramp}), we obtain
\bea \label{qfourfinal}
A^{(qq\tilde{q}\tilde{q})}=-\delta^{\rho}_{\;\;\lambda}\delta^{\sigma}_{\;\;\kappa}\frac{\LA 34\RA^2}{\LA 23\RA\LA 41\RA}
-\epsilon^{\rho\sigma\tau}
\epsilon_{\tau\kappa\lambda}\frac{\LA 34\RA}{\LA 12\RA}.
\eea
It is easy to see that this result is in agreement with the field theory predictions. 
An important remark is now in order. As we can also see in Figure 6, there are two types of contributions to
this scattering process. One of them comes from a Yukawa type interaction term while the other comes from the usual matter-gluon interaction.
In general these two interaction terms would be weighted with the appropriate independent coupling constant. It would be interesting to check if the consistency of the twistor method constraints the couplings to be in the conformal region of the moduli space.  Amplitudes like the one considered in this example might provide some insight on how to move away from the point in the moduli space where all the couplings are equal.


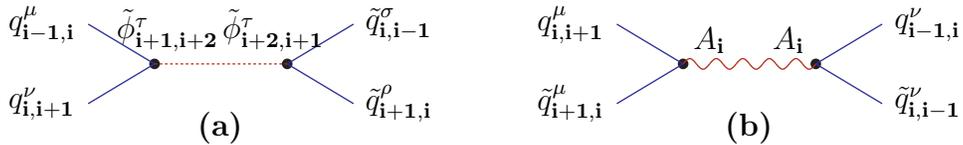
\begin{figure}[ht] \label{figampl2}
\begin{center}
\begin{picture}(300,50)(0,0)
\put(-30,0){
\Vertex(50,25){2} \Vertex(100,25){2}
\SetColor{BrickRed}
\DashLine(50,25)(100,25){1}
\SetColor{Blue}
\Line(25,10)(50,25)
\Line(25,40)(50,25)
\Line(100,25)(125,40)
\Line(100,25)(125,10)
\Text(20,40)[r]{$q^{\mu}_{{\bf i-1},\bi}$}\Text(20,10)[r]{$q^{\nu}_{\bi,{\bf i+1}}$}
\Text(130,10)[l]{$\tilde{q}^{\rho}_{{\bf i+1},\bi}$}\Text(130,40)[l]{$\tilde{q}^{\sigma}_{\bi,{\bf i-1}}$}
\Text(55,30)[b]{$\tilde{\phi}^{\tau}_{{\bf i+1},{\bf i+2}}$}\Text(95,30)[b]{$\tilde{\phi}^{\tau}_{{\bf i+2},{\bf i+1}}$}
\Text(75,1)[c]{\bf{(a)}}
}
\put(170,0){
\Vertex(50,25){2} \Vertex(100,25){2}
\SetColor{BrickRed}
\Photon(50,25)(100,25){2}{5}
\SetColor{Blue}
\Line(25,10)(50,25)
\Line(25,40)(50,25)
\Line(100,25)(125,40)
\Line(100,25)(125,10)
\Text(20,40)[r]{$q^{\mu}_{\bi,{\bf i+1}}$}\Text(20,10)[r]{$\tilde{q}^{\mu}_{{\bf i+1},\bi}$}
\Text(130,10)[l]{$\tilde{q}^{\nu}_{\bi,{\bf i-1}}$}\Text(130,40)[l]{$q^{\nu}_{{\bf i-1},\bi}$}
\Text(60,29)[b]{$A_{\bi}$}\Text(90,29)[b]{$A_{\bi}$}
\Text(75,1)[c]{\bf{(b)}}
}
\end{picture}
\caption{The two Feynman diagrams that contribute to tree-level $(q\,q\,\tilde{q}\,\tilde{q})$ scattering.}
\end{center}
\end{figure}

\subsection{An $\mathcal{N}=2$ orbifold}

\begin{figure}
\begin{center}
\includegraphics[width=75mm]{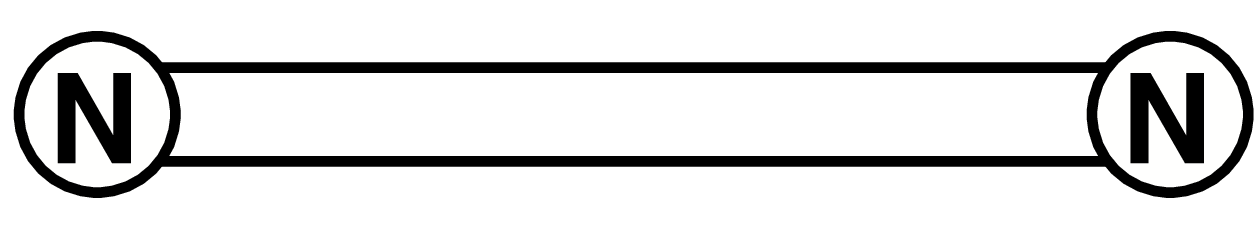}
\caption{The $\mathcal{N}=2$ quiver for $k=2$ and $a_A=(1,1,0,0)$.}\label{figure1}
\end{center}
\end{figure}

We now move on to the case in which $\Gamma\in Z_k \subset  SU(2)_R$. This breaks $SU(4)_R\rightarrow SU(2)_R$, thus giving $\mathcal{N}=2$. For simplicity, we investigate the particular choice of $k=2$ and $a_A=(1,1,0,0)$, see (\ref{psiorb}).
The fields surviving the orbifold projection are organized into two $\mathcal{N}=2$ vector multiplets and two hypermultiplets. The resulting gauge group is $U(N)\times U(N)$. The associated quiver diagram has two nodes and two links and is given in Figure \ref{figure1}. As in the previous $\mathcal{N}=1$ case, this theory is superconformal.

The field content of the gauge sector is 
\bea
A_{\bi},\;\;\;G_{\bi},\;\;\;\lambda_{\bi}^{a},\;\;\;\tilde{\lambda}_{\bi}^{a}, \;\;\; \phi^{m}_{\bi}
\eea
where $A_{\bi}\equiv A^{I\bi}_{\;\;J\bi}$, $G_{\bi}\equiv G^{I\bi}_{\;\;J\bi}$, $\lambda_{\bi}^{a}\equiv (\chi_{3,4})^{I\bi}_{\;\;J\bi}$, $\tilde{\lambda}_{\bi}^{a}\equiv (\tilde{\chi}_{3,4})^{I\bi}_{\;\;J\bi}$, and $\phi^{m}_{\bi}\equiv (\phi_{12,34})^{I\bi}_{\;\; J\bi}$. The nodes of the quiver are labelled by $\bi=1,2$. The index $a$ is the $SU(2)_R$ index, whereas $m$ labels the two real components of the complex scalar field.
The matter sector has two bifundamental hypermultiplets
\bea
q^{\mu}_{ \bi,{\bf i+1}}, \;\;\; \tilde{q}^{\mu}_{{\bf i+1},\bi}, \;\;\; H^{\mu}_{\bi,{\bf i+1}}, \;\;\;\tilde{H}^{\mu}_{{\bf i+1}, \bi}
\eea
with the index $\mu=1,2$ labels the hypermultiplets. The quarks $q^{\mu}_{ \bi,{\bf i+1}}$ are given by $(\chi^{1,2})^{I\bi}_{\;\;J,{\bf i+1}}$ and the anti-quarks $\tilde{q}^{\mu}_{{\bf i+1},\bi}$ by $(\tilde{\chi}^{1,2})^{I,{\bf i+1}}_{\;\;J\bi}$. The four scalars $H^{\mu}$ and $\tilde{H}^{\mu}$ come from $\phi_{13}, \phi_{14}, \phi_{23},$ and $\phi_{24}$.

The projected action and the D1-D5 interaction term can be obtained in similarly to the previous $\mathcal{N}=1$ case. 

\subsubsection{MHV Amplitudes for the $\mathcal{N}=2$ orbifold}

\paragraph{Example 1: Amplitudes like $(A\ldots A\,G\,G)$, $(A\ldots A\,G,\lambda\,\tilde{\lambda})$, and $(A\ldots A\, G\, q\,\tilde{q})$}

We consider the scattering  process between the following particles
$(A\,A\,G\,\lambda^a\,\tilde{\lambda}^c)$.
According to the twistor string method, we should compute
\bea\label{AAGlambdatlambda}
A^{(AAG\lambda\tilde{\lambda})}=\int d^8\theta \, \psi^1_3\psi^2_3\psi^3_3\psi^4_3\psi^{a}_4\psi^1_5\psi^2_5\psi^{b}_5
\epsilon_{bc}\frac{1}{\LA 12\RA\LA23\RA\LA34\RA\LA41\RA}.
\eea 
There are only two possible contractions between the different $\psi$'s, which yield
\bea
A^{(AAG\lambda\tilde{\lambda})}&=&(\delta^{3b}\delta^{4a}-\delta^{3a}\delta^{4b})\epsilon_{bc}
\frac{\LA35\RA^3\LA34\RA}{\LA 12\RA\LA23\RA\LA34\RA\LA45\RA\LA51\RA} \nonumber \\
&=&\delta^{ac}(\delta^{4c}-\delta^{3c})\frac{\LA35\RA^3}{\LA 12\RA\LA23\RA\LA45\RA\LA51\RA}.
\eea
As we immediately see, we recovered the familiar result. 

\paragraph{Example 2: Amplitude $(\lambda \,q\,\tilde{q}\,\tilde{\lambda})$} 

We will now apply the same method in order to compute the scattering amplitude $(\lambda^a \,q^{\mu}\,\tilde{q}^{\rho}\,\tilde{\lambda}^c)$ between a gluino--antigluino pair and a quark--antiquark one.
To this end, we need to calculate the following integral
\bea\label{lambdaq2amp}
A^{(\lambda q\tilde{q}\tilde{\lambda})}=\int d^8\theta \, \psi^{a}_1\psi{^\mu}_2(\psi^3_3\psi^4_3\psi^{\nu}_3)(\psi^1_4\psi^2_4\psi^{b}_4)
\epsilon_{\nu\rho}\epsilon_{bc}\frac{1}{\LA 12\RA\LA23\RA\LA34\RA\LA41\RA}
\eea 
where $a,b,c=3,4$ and $\mu,\nu,\rho=1,2$. 
To perform the integration we need to sum over all the possible contractions between the fermionic 
coordinates of supertwistor space. In this example we can split the fermions into two groups, with no contractions among fermions 
belonging to different groups. In each group, fermions can be contracted in two different ways
\bea \label{g1}
\begin{split}
(a)&\;
\int d^4\, \theta
\Wick{4mm}\psi^{a}_1\psi^{3}_3
\Wick{4mm}\psi^4_3\psi^{b}_4=\delta^{a3}\delta^{4b}\LA13\RA\LA34\RA\\
(b)&\;
\int d^4\, \theta
\Wick{10mm}\psi^{a}_1\Wickunder{10mm}\psi^3_3\psi^4_3\psi^{b}_4=-\delta^{a4}\delta^{b3}\LA13\RA\LA34\RA
\end{split}
\eea
and
\bea \label{g2}
\begin{split}
(a)&\;
\int d^4\, \theta
\Wick{14.5mm}\psi^{\mu}_2\Wickunder{4mm}\psi^{\nu}_3\psi^1_4\psi^2_4=\delta^{\mu2}\delta^{\nu1}\LA24\RA\LA34\RA\\
(b)&\;
\int d^4\, \theta
\Wick{10mm}\psi^{\mu}_2\Wickunder{10mm}\psi^{\nu}_3\psi^1_4\psi^2_4=-\delta^{\mu1}\delta^{\nu2}\LA24\RA\LA34\RA.
\end{split}
\eea
We then substitute (\ref{g1}) and (\ref{g2}) into (\ref{lambdaq2amp}) and use the Schouten Identity to obtain
\bea
A^{(\lambda q\tilde{q}\tilde{\lambda})}=\delta^{ac}\delta^{\mu\rho}(\delta^{\mu 2}-\delta^{\mu 1})(\delta^{a3}-\delta^{a4})
\left(\frac{{\LA34\RA}^2}{\LA23\RA\LA41\RA}-\frac{\LA34\RA}{\LA12\RA}\right).
\eea
We see from the form of the result that there are exactly two distinct spinor product structures. They reflect contributions to the scattering 
process from two different types of interactions: the former is the standard quark-gluon interaction and the latter is of Yukawa type. Figure
8 shows the corresponding Feynman diagramms. This is in accordance with the usual field theory calculations.  

Other amplitudes, with quarks or scalars as external particles, can be computed in a similar fashion. They usually retain the feature of receiving
contributions from multiple interaction processes/vertices. 


\begin{figure}[ht] \label{figampl0}
\begin{center}
\begin{picture}(300,50)(0,0)
\put(-30,0){
\Vertex(50,25){2} \Vertex(100,25){2}
\SetColor{BrickRed}
\DashLine(50,25)(100,25){1}
\SetColor{Blue}
\Line(25,10)(50,25)
\Line(25,40)(50,25)
\Line(100,25)(125,40)
\Line(100,25)(125,10)
\Text(20,40)[r]{$\lambda^{b}_{\bi}$}\Text(20,10)[r]{$q^{\mu}_{\bi,{\bf i+1}}$}
\Text(130,10)[l]{$\tilde{q}^{\rho}_{{\bf i+1},\bi}$}\Text(130,40)[l]{$\tilde{\lambda}_{\bi}$}
\Text(60,27)[b]{$\tilde{H}^{\nu\, a}_{{\bf i+1},\bi}$}\Text(90,27)[b]{$H^{\nu\, a}_{\bi,{\bf i+1}}$}
\Text(75,1)[c]{\bf{(a)}}
}
\put(170,0){
\Vertex(50,25){2} \Vertex(100,25){2}
\SetColor{BrickRed}
\Photon(50,25)(100,25){2}{5}
\SetColor{Blue}
\Line(25,10)(50,25)
\Line(25,40)(50,25)
\Line(100,25)(125,40)
\Line(100,25)(125,10)
\Text(20,40)[r]{$\tilde{\lambda}_{\bi}$}\Text(20,10)[r]{$\lambda_{\bi}$}
\Text(130,10)[l]{$q^{\mu}_{\bi,{\bf i+1}}$}\Text(130,40)[l]{$\tilde{q}^{\rho}_{{\bf i+1},\bi}$}
\Text(60,29)[b]{$A_{\bi}$}\Text(90,29)[b]{$A_{\bi}$}
\Text(75,1)[c]{\bf{(b)}}
}
\end{picture}
\caption{The two Feynman diagrams that contribute to tree-level $(\lambda\, q\,\tilde{q}\,\tilde{\lambda})$ scattering.}
\end{center}
\end{figure}
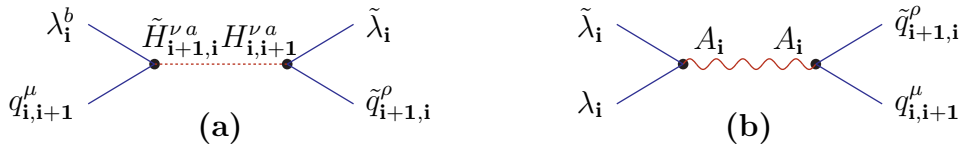

\section{Conclusion}
In this paper we have investigated $Z_k$ fermionic orbifolds of the topological B-model on $CP^{(3|4)}$ to reduce the amount of supersymmetry of the dual $\mathcal{N}=4$ Super Yang-Mills theory. This is possible since the Fermi-Fermi part of the $PSU(4|4)$ isometry group of $CP^{(3|4)}$ is precisely the $SU(4)_R$ $R$-symmetry. We have discussed how the projection acts on both the D5 and the D1 branes.
As examples we have considered $\mathcal{N}=1$ and $\mathcal{N}=2$ orbifolds and obtained the corresponding quiver theories. Several amplitudes have been computed and shown to agree with the field theory results. 

Throughout the paper we have worked only at the point in the moduli space where all gauge couplings constants are equal. It would be interesting to study the full moduli space of superconformal couplings and understand its interpretation in twistor string theory. Some indications on the origin of these moduli have been given in discussing the action of the orbifold on the D1 brane sector. Since these moduli are usually interpreted as coming from twisted fields it would be worth studying the closed string sector and identify the twisted states. The fermionic orbifold does not have an obvious geometrical meaning. The study of the twisted sector may be useful to shed some light on the geometrical interpretation of the orbifold.

\paragraph{Acknowledgements}
It is a pleasure to thank Martin Ro\v cek and Albert Schwarz for useful discussions and comments. We would also like to express our gratitude to the participants and the organizers of the second Simons Workshop in Mathematics and Physics at Stony Brook for creating a stimulating environment from which we benefited greatly. We acknowledge partial financial support through NSF award PHY-0354776.

\paragraph{Note added in proof}
After completion of our computations, while writing up the results, we became aware of \cite{rey}, which has overlap with our work. 


\end{document}